\documentclass{webofc}
\usepackage[varg]{txfonts}   

\begin{document}
\begin{flushright}
INR-TH-2018-020
\end{flushright}

\title{On the relation between  pole and running heavy quark  masses beyond the four-loop approximation}

\author{\firstname{A. L.} \lastname{Kataev}\inst{1,2}\fnsep\thanks{\email{kataev@ms2.inr.ac.ru}} \and
        \firstname{V. S.} \lastname{Molokoedov}\inst{1,2,3}\fnsep\thanks{\email{viktor_molokoedov@mail.ru}} 
        }

\institute{Institute for Nuclear
Research of the Russian Academy of Sciences, \\60th October Anniversary prospect 7a, Moscow, 117312, Russia
\and
           Moscow Institute of Physics and Technology,\\Institusky per. 9, Dolgoprudny, Moscow region, 141700, Russia
\and
           Landau Institute for Theoretical Physics of the Academy of Sciences of Russia,\\Akademika Semenova avenue 1-a, Chernogolovka, Moscow Region, 142432, Russia}

\abstract{The effective charges motivated method 
is applied to the relation between pole and  $\rm{\overline{MS}}$-scheme  heavy quark masses
to  study  high order perturbative QCD corrections in  the observable quantities   proportional 
to the  running quark masses.  The 
 non-calculated  five-  and six-loop perturbative  QCD coefficients are estimated.
 This  approach  predicts  for these terms  the sign-alternating expansion  in   powers  of number  of lighter 
flavors $n_l$, while the analyzed recently infrared renormalon  
 asymptotic expressions   do not reproduce the same  behavior.  
We  emphasize  that coefficients of the  quark mass relation     contain  proportional to $\pi^2$   effects, which result from analytical    continuation from the Euclidean region, where the   scales of the running masses  and QCD coupling constant are   initially  fixed,  to the Minkowskian region,   where the pole masses and the running QCD parameters  
 are determined.  For the  $t$-quark  the   asymptotic  nature  of the non-resummed  PT  mass relation  does not manifest itself at six-loops, while  for the $b$-quark the minimal  PT term appears   at the    probed by   direct  calculations four-loop level. The  recent  infrared   renormalon 
 based studies support these conclusions.}

\maketitle

\section{Introduction}
\label{intro}
It is well-known that the masses of charm,  bottom and top-quarks are one of the most important QCD parameters, which are  relevant for processing different  data, obtained at LHC and Tevatron. The  pole and $\rm{\overline{MS}}$-scheme running heavy quark  masses are the 
 generally accepted definitions for  these parameters.  The first ones  are determined by the position of  pole of the renormalized  fermion  propagator  at Minkowskian region  $k^2=M^2_q$. The scale-dependence of the  second ones  are 
defined from the solution of the following renormalization-group equation:
\begin{equation}
\label{ms-5}
\frac{\overline{m}_q(s)}{\overline{m}_q(\mu^2)}={\rm{exp}}\bigg(\int\limits_{a_s(\mu^2)}^{a_s(s)}  \frac{\gamma_m(x)}{\beta (x)}~dx\bigg)~.
\end{equation}
Here  $\overline{m}_q(\mu^2)$ are the running masses of  heavy quarks,  normalized at the  Minkowskian  scale $\mu^2$, $a_s(\mu^2)=\alpha_s(\mu^2)/\pi$ is the renormalized QCD coupling constant in the $\rm{\overline{MS}}$-scheme, $s$ is the energy time-like variable. The RG $\beta$-function and anomalous mass dimension $\gamma_m$ are defined as:
\begin{eqnarray}
\label{RG-eq}
\mu^2\frac{\partial a_s}{\partial \mu^2}=\beta(a_s)=-\sum\limits_{n=0}^{\infty} \beta_n a^{n+2}_s~, ~~~~~~~ \mu^2\frac{\partial \log(\overline{m}_q)}{\partial \mu^2}=\gamma_m(a_s)=-\sum\limits_{n=0}^{\infty}\gamma_n a^{n+1}_s~.
\end{eqnarray}
Their five-loop approximations in the $\rm{\overline{MS}}$-scheme can be found in works of \cite{Baikov:2016tgj, Herzog:2017ohr} and \cite{Baikov:2014qja} correspondingly. It is worth to emphasize   that the renormalization scale  $\mu^2$ 
in Eqs.(\ref{ms-5}, \ref{RG-eq}) may be  initially defined in the Euclidean   region. 
As was shown in \cite{Bigi:1994em, Beneke:1994sw, Beneke:1994rs,Beneke:1994qe,Ball:1995ni, Beneke:1998ui}
the pole quark mass relation  is  sensitive to the long-distance infrared   renormalon (IRR)  effects. This IRR 
sensitivity leads to the $\mathcal{O}(\Lambda_{{\rm{QCD}}})$ renormalon ambiguity in  the determination of  heavy quark masses, which is related to the first $\delta$=1/2 IRR  pole in the $\delta$-plane of  the  Borel transform for the  PT QCD relation between pole and running   heavy quark masses  \cite{Beneke:1994sw}. 
Therefore,  the significant attention  is paid to
the determinations  of  masses in the $\rm{\overline{MS}}$-scheme, which do not  imply the need for knowledge of the long-distance contributions.  Indeed, their definition within dimensional regularization presumes  taking  into account of  the UV divergent poles only. Therefore sometimes the running $\rm{\overline{MS}}$ heavy quark masses are called the short-distance masses, which unlike  pole masses have nothing in common with  the  IRR  effects.

In view of all mentioned above it is of interest to consider the relation between   pole and running masses of heavy quarks: 
\begin{equation}
\label{t^M_n}
M_q=\overline{m}_q(\overline{m}^2_q)\sum\limits_{n=0}^{\infty} t^M_n a^n_s(\overline{m}^2_q)~.
\end{equation}
The choice  $\mu^2=\overline{m}^2_q$ corresponds to the 
commonly  accepted way of  fixation of the renormalization scale in the  Minkowskian region,  wherein coefficients $t^M_n$ are polynomials in powers of the number of massless flavors $n_l$
\begin{equation}
\label{t-n-l}
t^M_n=\sum\limits_{k=0}^{n-1} t^M_{nk} n^k_l~
\end{equation}
with the initial condition $t^M_0=1$. 
Note that we  consider the approximation when  heavy flavor number $n_f$  of active quarks  
are defined as $n_f=n_l+1$. 
 The results of calculations of  one, two and three-loop corrections  to (\ref{t^M_n}) were obtained  in \cite{Tarrach:1980up}, \cite{Gray:1990yh, Avdeev:1997sz, Fleischer:1998dw}, \cite{Melnikov:2000qh, Chetyrkin:1999qi} respectively. The four-loop contributions $t^M_{43}$ and $t^M_{42}$ are known from the analytical calculations, performed in \cite{Lee:2013sx}. The rest $\mathcal{O}(a^4_s)$-corrections, namely $t^M_{41}$ and $t^M_{40}$-terms, are not yet computed in analytical form, but are evaluated numerically with the corresponding theoretical uncertainties. The total  expressions   for the  $t^M_4$-term at fixed numbers of massless flavors  $n_l=3, 4, 5$
 were evaluated   in \cite{Marquard:2015qpa}. Using these numerical  results, supplemented with the analytical 
information on $t^M_{43}$ and $t^M_{42}$-terms, the authors of  Ref.\cite{Kiyo:2015ooa} 
presented the first estimates of the unknown previously $t^M_{41}$ and $t^M_{40}$-coefficients 
by means of the renormalon-inspired approach. Independently, the numerical values of these terms were determined 
in \cite{Kataev:2015gvt, Kataev:2016jai}  with the help of mathematically self-consistent  least squares  (LS) method, which is 
well-defined  procedure of solving the overdetermined system of linear equations with the fixation  of theoretical 
inaccuracies of the  central values of the  obtained results.   
 Recently the updated results of  the numerical  evaluation of $t^M_4$-terms  were obtained by means of the Monte-Carlo methods in \cite{Marquard:2016dcn} not for three values of $n_l$ only but  at  extra 18 values as well, which corresponds 
to the additional  points in the studied interval  $0\leq  n_l \leq 20$.  The results 
of these updated computations were  obtained with  considerably  smaller
 uncertainties than the ones, presented in \cite{Marquard:2015qpa}.
 This  prompted us  to reconsider  the results of \cite{Kataev:2015gvt} and to get  new values of 
 $t^M_{41}$, $t^M_{40}$-terms and their uncertainties using the same  LS method \cite{Kataev:2018gle}.
The  central values of new results have changed slightly (see the Note added to \cite{Kataev:2015gvt}
as well).  This is  related to 
slight change of the central values of the  improved more precise  numbers  from  Ref.\cite{Marquard:2016dcn}. 
Moreover,  thanks to the  increase of the number  of analyzed 
  by LS  method  equations, fixed by the results 
from \cite{Marquard:2016dcn}, which 
have  smaller numerical  errors than the outcomes  of previous  calculations  \cite{Marquard:2015qpa}, the obtained in \cite{Kataev:2018gle} solutions of larger  system of linear  equations turned out to be drastically   more precise than the ones, obtained previously in Ref.\cite{Kataev:2015gvt}.
This feature is in agreement with general property of the mathematical LS approach.  
Leaving the discussions  of the  technical issues of this method aside we will return now to the  consideration  of the asymptotic  structure of Eq.(\ref{t^M_n}).

Due to the manifestation of the IRRs   in the Borel image for the pole-running heavy quark mass relation, it is possible to 
conclude  that the mass conversion formula of Eq.(\ref{t^M_n}) is asymptotic one with the sign-constant factorially  growing 
high order PT coefficients  $t^M_n$. This means that at some  orders of PT the series in (\ref{t^M_n}) start to diverge. Indeed,  from the results of direct  calculations, performed  at the two-loop level in the works  \cite{Gray:1990yh, Avdeev:1997sz, Fleischer:1998dw} 
and  at the three-loop level in Refs.\cite{Melnikov:2000qh, Chetyrkin:1999qi} it is possible to conclude that the perturbative QCD relation between pole and running  charm-quark mass  diverges from the second (or the third) order of PT. The situation with the mass conversion formula for the  $b$-quark is more delicate. Indeed, 
 its PT  high-order contributions   decrease  up to four-loop level, although quite slowly. The four-loop contribution,  numerically 
evaluated  in \cite{Marquard:2015qpa, Marquard:2016dcn}, is very close to its three-loop term. This allows to affirm that the renormalon nature of the PT series for the $\rm{\overline{MS}}$-on-shell mass relation for bottom-quark is manifesting itself from the $\mathcal{O}(a^4_s)$  contribution. However, in the case of the  pole-running  top-quark mass relation the perturbative outburst of the corresponding series is not manifesting  itself at the four-loop level. In order to understand  when the truncated perturbative series can be still used  for case of 
the $t$-quark  pole mass, it is necessary to estimate  high-order corrections to its relation. This problem is analyzed  by us below.
 
\section{The effective-charges motivated method: from 
 the Euclidean to Minkowski region}
\label{Euclidean}

Let us now apply the used  in \cite{Chetyrkin:1997wm} approach for estimations of the high-order PT QCD corrections to  the relation between different definitions of  heavy quarks. This approach is following the lines of the developed in \cite{Kataev:1995vh} method of probing the values of high-order perturbative corrections to the renormalization-group invariant quantities, which in its turn is  based on the  concepts of the effective-charges (ECH) 
of Ref.\cite{Grunberg:1982fw}. The considered in \cite{Chetyrkin:1997wm} approach
 was already used in \cite{Kataev:2010zh}  for estimating four-loop PT QCD  corrections 
to the  expression between pole and $\rm{\overline{MS}}$ running heavy quark masses, numerically evaluated 
later on in \cite{Marquard:2015qpa}.  As was shown in Refs.\cite{Chetyrkin:1997wm, Kataev:1995vh}, 
from general grounds it is more theoretically justified to use the ECH-motivated procedure to the physical quantities, defined  in the Euclidean space-time region, and then, if necessary,  to translate the expressions of the corresponding  PT corrections to the Minkowskian region. This leads to the appearance of the   proportional to powers of  $\pi^2$-terms in the coefficients of the PT series, which relate the  quantities, defined in the Minkowskian region. In our case these quantities are the pole and running heavy quark masses.

To clarify how this procedure is working 
we start from the following formal dispersion representation for the  pole masses of the heavy quarks, first considered in \cite{Chetyrkin:1997wm}:
\begin{equation}
\label{Massdispersion}
M_q=\frac{1}{2\pi i}\int_{-\overline{m}_q(\overline{m}^2_q)-i\epsilon}^{-\overline{m}_q(\overline{m}^2_q)+i\epsilon}
ds^{\prime}\int_0^{\infty}\frac{T(s)}{(s+s^{\prime})^2}ds~.
\end{equation}
The function $T(s)$  has the meaning of the spectral density and it is defined as $T(s)=\overline{m}_q(s)\sum_{n=0}^{\infty} t^M_n a^n_s(s)$, where coefficients $t^M_n$ coincide with the ones, which  enter  Eqs.(\ref{t^M_n}) and (\ref{t-n-l}). The expression (\ref{Massdispersion}) is similar  to  relation between the  determined in the Euclidean region 
  the $e^+e^-$ annihilation  Adler function and the Minkowskian time-like ratio $R(s)$, which is one of the  main characteristics
of the $e^+e^-$ annihilation  to hadrons process. 
The used  in  Ref.\cite{Chetyrkin:1997wm} dispersion relation of (\ref{Massdispersion}) leads to the conclusion that the
 analogue of the Adler function, namely the Euclidean  quantity $F(Q^2)$  may be defined as  
\begin{equation}
\label{dispersion}
F(Q^2)=Q^2\int\limits_{0}^{\infty} ds \frac{T(s)}{(s+Q^2)^2}~~~.
\end{equation}
 Within PT   QCD it is expressed through  the following   series  
\begin{equation}
\label{dispersion-1}
F(Q^2)=\overline{m}_q(Q^2)\sum\limits_{n=0}^{\infty}f^E_n a^n_s(Q^2)~.
\end{equation}
Taking into account the scale dependence of the $\rm{\overline{MS}}$-scheme coupling constant  and of 
 the running heavy quark masses 
and using expansions (\ref{Massdispersion}-\ref{dispersion-1}), we can fix the relations between  the 
 coefficients of the Minkowskian series  $t^M_n$ 
  and the determined in the  the Euclidean region    terms  $f^E_n$ as:
\begin{equation}
\label{f-t-D}
f^E_n=t^M_n+\Delta_n~. 
\end{equation}
In  accordance with equation (\ref{dispersion}) $\Delta_n$-contributions  contain the proportional to  $\pi^2$ effects of the analytical continuation. Their detailed derivation  at the six-loop level  is given  in \cite{Kataev:2018gle}.
Here we present the  numerical expressions of these contributions for the case of $SU(3)$ color gauge group only. They have the following form:
\begin{align}
\label{C2}
\Delta_0&=0~, ~~~~~~~~ \Delta_1=0~, \\ \nonumber
\Delta_2&=5.89434-0.274156n_l~, \\ \nonumber
\Delta_3&=105.6221-10.04477n_l+0.198002n^2_l~, \\
 \nonumber
\Delta_4&=2272.002-403.9489n_l+20.67673n^2_l-0.315898n^3_l~, \\
 \nonumber
\Delta_5&=56304.639-13767.2725n_l+1137.17794n^2_l-37.745285n^3_l+0.427523n^4_l~,  \\
 \nonumber
\Delta_6&=1633115.62\pm 347.65+(-518511.694\pm 56.723)n_l+(61128.1666\pm 4.7791)n^2_l \\ \nonumber
&+(-3345.0818\pm 0.1371)n^3_l+85.37937n^4_l-0.818446n^5_l.
\end{align}
The  analytical expressions for $\Delta_0-\Delta_4$-terms can be found in \cite{Chetyrkin:1997wm, Kataev:2010zh},
whereas the analytical expressions for $\Delta_5$ and $\Delta_6$ were obtained in \cite{Kataev:2018gle}. The 
 uncertainties,  which enter in the numerical expression for  $\Delta_6$-term  are determined by the corresponding inaccuracies of the four-loop numerical  $t^M_{41}$ and $t^M_{40}$-contributions to Eq.(\ref{t-n-l}), fixed  in \cite{Kataev:2018gle} by  the LS  method. 
Using   Eqs.(\ref{C2}) one can find   that the numerical values of the  $\Delta_n$-terms increase considerably  with the growth of the order $n$ of PT.  

At the next stage of application of  the ECH-motivated approach we fix  the effective charge $a^{eff}_s(Q^2)$ for the introduced in (\ref{dispersion-1}) 
 Euclidean quantity $F(Q^2)/\overline{m}_q(Q^2)$ 
\begin{equation}
\label{effective}
\frac{F(Q^2)}{\overline{m}_q(Q^2)}=f^E_0+f^E_1 a^{eff}_s(Q^2)~, ~~~~~~ a^{eff}_s(Q^2)=a_s(Q^2)+\sum\limits_{n=2}^{\infty}\phi_n a^n_s(Q^2)~,
\end{equation}
 where $\phi_n=f^E_n/f^E_1$. The coefficients of the ECH $\beta$-function, which is defined as $\beta^{eff}(a^{eff}_s)=-\sum_{n\geq 0}\beta^{eff}_n(a^{eff}_s)^{n+2}$, 
 are related to the 
coefficients $\beta_n$ determined in the  $\rm{\overline{MS}}$-scheme $\beta$-function of Eq.(\ref{RG-eq}) by the following renormalization-scheme invariant equations:
\begin{subequations}
\begin{eqnarray}
\label{betaef0-2}
\beta^{eff}_0&=&\beta_0~, ~~~~~~\beta^{eff}_1=\beta_1~, ~~~~~~\beta^{eff}_2=\beta_2-\phi_2\beta_1+(\phi_3-\phi^2_2)\beta_0~, \\
\label{betaef3}
\beta^{eff}_3&=&\beta_3-2\phi_2\beta_2+\phi^2_2\beta_1+(2\phi_4-6\phi_2\phi_3+4\phi^3_2)\beta_0~, \\
\label{betaef4}
\beta^{eff}_4&=&\beta_4-3\phi_2\beta_3+(4\phi^2_2-\phi_3)\beta_2+(\phi_4-2\phi_2\phi_3)\beta_1 \\ \nonumber
&+&(3\phi_5-12\phi_2\phi_4-5\phi^2_3+28\phi^2_2\phi_3-14\phi^4_2)\beta_0~, \\
\label{betaef5}
\beta^{eff}_5&=&\beta_5-4\phi_2\beta_4+(8\phi^2_2-2\phi_3)\beta_3+(4\phi_2\phi_3-8\phi^3_2)\beta_2 \\ \nonumber
&+&(2\phi_5-8\phi_2\phi_4+16\phi^2_2\phi_3-3\phi^2_3-6\phi^4_2)\beta_1 \\ \nonumber
&+&(4\phi_6-20\phi_2\phi_5-16\phi_3\phi_4+48\phi_2\phi^2_3
-120\phi^3_2\phi_3+56\phi^2_2\phi_4+48\phi^5_2)\beta_0~.
\end{eqnarray} 
\end{subequations}
The   starting point   of the  ECH-motivated estimating procedure of Ref.\cite{Kataev:1995vh} is the ansatz    $\beta^{eff}_n=\beta_n$,  which should be applied separately   at each  order of PT beginning  from the three-loop one.  
 In the  case  of the QCD  relation between different definitions of heavy quark masses it was 
used  at the three-loop level
 in Refs.\cite{Chetyrkin:1997wm, Kataev:2010zh} and allowed to get the estimates for the coefficients   $\phi_3=f^E_3/f^E_1$  from Eq.(\ref{betaef0-2}). 
 Further  application of the 
 relation (\ref{f-t-D})  with  the given in  Eq.(\ref{C2})
 numerical 
expressions for  the  typical to the Minkowski 
region term   $\Delta_3$ leads to good agreement of  
 the obtained in  \cite{Chetyrkin:1997wm,  Kataev:2010zh} 
 approximate  expression  for the 
 $t^M_3$-coefficient   (we denote it as  $t^{M, \; ECH}_3$) 
 with the explicit  three-loop result, obtained   in  \cite{Melnikov:2000qh, Chetyrkin:1999qi}. 
It turned out later  that the estimated in Ref.\cite{Kataev:2010zh} by the similar way  values of 
 the coefficient $t^{M, \; ECH}_4$ at $n_l=3,4,5$ 
are  also  in reasonable  agreement with the results of the numerical calculations, performed in \cite{Marquard:2015qpa}
(see the work of Ref.\cite{Marquard:2016dcn} as well).  

These facts  serve a-posteriori arguments in favor  of the applicability of this ECH-inspired method, 
supplemented with the  explicit expressions  for  the proportional to $\pi^2$  effects of analytical continuation,  
at higher orders of PT as well.  In  Ref.\cite{Kataev:2018gle}, which is summarized in brief here, we applied the conditions 
$\beta^{eff}_4=\beta_4$ and $\beta^{eff}_5=\beta_5$ for Eqs.(\ref{betaef4}) and (\ref{betaef4}) 
and got the numerical  estimates for the  coefficients $t^{M, \; ECH}_5$ and $t^{M, \; ECH}_6$ after taking into account 
the  analytical continuation contributions $\Delta_5$ and $\Delta_6$. 
The concrete numerical results are presented and discussed in Sec.5. Here we only note, that 
while getting the estimate for the coefficient  $t^M_6$  from Eqs.(\ref{betaef5}) and (\ref{f-t-D}) 
 in addition to the theoretical  
ansatz  $\beta^{eff}_5=\beta_5$, which of  course has definite unfixed theoretical  uncertainties, 
 the obtained by the ECH-motivated method at the five-loop level  estimate for $f^{E}_5$-coefficient was used. Its application  at the  six-loop order  leads to the  additional  theoretical ambiguity of the   estimated  value   
of the $\mathcal{O}(a^6_s)$-correction to the pole-running heavy quark masses relation, which is not possible to fix.
 However, in order to  study   whether there may be extra theoretical ambiguities in  the results of applications of 
the widely spreaded  IRR approach, recently used in  \cite{Beneke:2016cbu} to analyze the uncertainties of the asymptotic QCD predictions for the coefficients of heavy quark masses relation,  we will compare in Sec.5 its  outcomes  with 
the five and six-loop estimates of the same terms,  obtained by means not related to the IRR-approach   ECH-inspired methods.

\section{The effective-charges motivated method: direct 
application in the Minkowski region}
\label{Minkowski}

Since the pole masses of heavy quarks are defined in the Minkowski region, 
it is also  worth   to consider the predictions of the coefficients in  the pole-running heavy quark mass relation applying the  ECH-motivated approach of \cite{Kataev:1995vh}   in the time-like region directly. 
This was first done in \cite{Chetyrkin:1997wm} (see \cite{ Kataev:2010zh} as well) 
constructing   the Minkowskian analogs of Eqs.(\ref{effective}-\ref{betaef3}) for  the  
spectral function of  Eq.(\ref{Massdispersion})  
\begin{equation}
\label{Mdefinition}
T(s)=\overline{m}_q(s)\sum\limits_{n=0}^{\infty}t^M_n a^n_s(s)~.
\end{equation}
Defining the  ECH  $a^{eff}_s(s)$ for the  quantity $T(s)/\overline{m}_q(s)$ and   the corresponding   
$\tilde{\beta}^{eff}(a^{eff}_s)$-function with 
 coefficients $\tilde{\beta_n}^{eff}$,  fixed by the  replacements  $\phi_n\rightarrow\phi^M_n=t^M_n/t^M_1$ in Eqs.(\ref{effective}-\ref{betaef5}),  one can get the estimates for coefficients $t_n^{M, \; ECH\;direct}$
 of pole-running heavy quark mass relation directly in the Minkowskian region 
after using the ansatz $\tilde{\beta}^{eff}_n=\beta_n$ at  $n\geq 2$.   In Ref.\cite{Kataev:2010zh} it was noticed that  the  numerical expressions of the third and fourth  coefficients $t^{M, \; ECH\;direct}_3$ and $t^{M, \; ECH\;direct}_4$
 of the relations between pole and running charm, 
bottom and top-quark masses  are in satisfactory agreement with the  values  of $t^{M, \; ECH}_3$ and $t^{M, \; ECH}_4$-terms, obtained  as  described above in Sec.2
(for the detailed comparison see  \cite{Kataev:2018gle}). The   estimated  five- and 
six-loop terms, obtained in the Minkowski region directly,    have the following form 
\begin{subequations}
\begin{align}
\label{t^ECHd_5}
t^{M, \; ECH\;direct}_5
&\approx \frac{1}{3\beta_0(t^M_1)^3}\bigg[3t^M_2(t^M_1)^3\beta_3+
t^M_3(t^M_1)^3\beta_2-4(t^M_2t^M_1)^2\beta_2 \\ \nonumber
&+2t^M_3t^M_2(t^M_1)^2\beta_1-t^M_4(t^M_1)^3\beta_1
+12t^M_4t^M_2(t^M_1)^2\beta_0+5(t^M_3t^M_1)^2\beta_0
\\  \nonumber 
&+14(t^M_2)^4\beta_0-28t^M_3(t^M_2)^2t^M_1\beta_0\bigg]
 \\ 
\label{t^ECHd_6} 
t^{M, \; ECH\;direct}_6 &\approx \frac{1}{12\beta^2_0(t^M_1)^4}\bigg[48 t^M_4 t^M_3 (t^M_1)^3 \beta_0^2+72t^M_4(t^M_1t^M_2)^2\beta_0^2+12t^M_2(t^M_1)^4\beta_0\beta_4 
\\ \nonumber
&+ 136(t^M_2)^5\beta_0^2-200t^M_3t^M_1(t^M_2)^3\beta_0^2-20t^M_4t^M_2(t^M_1)^3\beta_0\beta_1 -(t^M_1)^3(t^M_3)^2\beta_0\beta_1 \\ \nonumber
&+48t^M_3(t^M_1t^M_2)^2\beta_0\beta_1 
-10t^M_1(t^M_2)^4\beta_0\beta_1 
-44t^M_2(t^M_1 t^M_3)^2\beta_0^2
+6t^M_3(t^M_1)^4\beta_0\beta_3 \\ \nonumber
&+ 36(t^M_1)^3(t^M_2)^2\beta_0\beta_3
 -56(t^M_1)^2(t^M_2)^3\beta_0\beta_2 + 2t^M_4(t^M_1)^4\beta^2_1 - 4t^M_3t^M_2(t^M_1)^3\beta^2_1 \\  \nonumber
&+ 8t^M_3t^M_2(t^M_1)^3\beta_0\beta_2
- 6t^M_2(t^M_1)^4\beta_1\beta_3 - 2t^M_3(t^M_1)^4\beta_1\beta_2 + 8(t^M_1)^3(t^M_2)^2\beta_1\beta_2\bigg]
\end{align}
\end{subequations} 
and like the two-,  three- and four-loop analogs  can be expressed  as 
\begin{equation}
\label{CoeffMinkoskian}
t^{M, \; ECH\;direct}_n= f_n^{E}-\tilde{\Delta}_n~~~.
\end{equation}
The made in Refs.\cite{Kataev:2010zh, Kataev:2018gle}
  observations that $t^{M, \; ECH\;direct}_3\approx t^{M, \; ECH}_3$ and  $t^{M, \; ECH\;direct}_4\approx t^{M, \; ECH}_4$
 mean that $\Delta_{3}\approx \tilde{\Delta}_{3}$ and $\Delta_{4}\approx \tilde{\Delta}_{4}$,
 where terms in the l.h.s.   are the  exactly calculable analytical continuation effects. The presented in Sec.5 comparison of five- and six-loop estimates, obtained within 
both approaches,  demonstrates that the above mentioned approximate equality remains true at the 5 and 6-loops  
level as well. This demonstrates compatibility  of the estimates for heavy quark mass relations coefficients, obtained by applying 
ECH-inspired procedure both in the Euclidean and Minkowskian regions.  

\section{The estimates by the IRR-based approach}
\label{Renormalon}
The most wide-spreaded modern approach   of the analysis  of high-order PT QCD corrections to physical quantities is based on application 
of the renormalon technique (for the previous developments see  e.g. Refs.\cite{Beneke:1994sw, Beneke:1994rs,
Beneke:1994qe,Ball:1995ni, Beneke:1998ui}, \cite{Beneke:1992ea, Broadhurst:2000yc, Kataev:2001hy}). 
It  is related  in part  to  large $\beta_0$-expansion
(for the application of the latter one see e.g. \cite{Beneke:1994qe,Ball:1995ni, Broadhurst:2000yc, Kataev:2001hy}).
  As was already mentioned above the asymptotic structure of the  
PT QCD expression of the  pole heavy quark masses through the  $\rm{\overline{MS}}$-scheme  running ones  is governed  by the leading  IRR  contribution
\cite{Bigi:1994em, Beneke:1994sw}, 
 which makes the coefficients of this relation growing  factorially  with the 
increase  of  order of PT. Therefore it is important  to 
analyze  the region of applicability of the corresponding asymptotic PT series.
 For this aim we consider  the IRR-based  formula \cite{Beneke:1994rs}, which  predicts  the following factorial  behavior of the coefficients  $t^M_n$:
\begin{align}
\label{renormalon-dom}
t^{M, \; r-n}_n\xrightarrow{n\rightarrow\infty}\pi N_m (2\beta_0)^{n-1}\frac{\Gamma(n+b)}{\Gamma(1+b)}\bigg(1+
\sum\limits_{k=1}^3\frac{s_k}{(n+b-1)\dots (n+b-k)}
+\mathcal{O}(n^{-4})\bigg)~,
\end{align}
where $\Gamma(x)$ is the Euler Gamma-function, $b=\beta_1/(2\beta^2_0)$ and  the values of the sub-leading coefficients $s_k$ can be found in   \cite{Pineda:2001zq, Beneke:2016cbu}. Note that our notations and normalizations differ from those introduced in Refs.\cite{Beneke:1994sw, Beneke:1994rs,  Beneke:2016cbu}. In the presented below discussions 
the coefficients of the RG $\beta$-function depend on $(n_l-1)$ numbers of  flavors.

The normalization factor $N_m$ in  Eq.(\ref{renormalon-dom}) is the function of 
 $n_l$ and of the order $n$ of PT. Unfortunately, its explicit form is not known. 
Moreover, the way of fixation of $N_m$-values is different in various works on the subject (see e.g. \cite{Campanario:2003ix, Ayala:2014yxa, Ayala:2016sdn}
and the  detailed work \cite{Pineda:2001zq}).
This fact introduces 
the important uncertainty  in   the IRR-based analysis. In our analysis  we use the 
given in Table 1  numerical results for $n_l$-dependence of  $N_m$, obtained in the process of four-loop analysis of  Ref.\cite{Beneke:2016cbu}\footnote{Note that in \cite{Kataev:2018gle} while applying the  asymptotic 
expression of Eq.(\ref{renormalon-dom}) for estimates of the five and six-loop corrections to the
$\rm{\overline{MS}}$-on-shell mass relation  for charm, bottom and top-quarks  
 the same approximate   average value  $N_m \approx 0.5$ was fixed, which in fact is too far from the result, obtained in Refs. \cite{Campanario:2003ix,
Pineda:2001zq}.}:
\begin{table}[!h]
\centering
\caption{The $n_l$  dependence of $N_m$ at the  the  fourth order of  PT.}
\label{Tab-1}
\begin{tabular}{lllllll}
\hline 
$~~n_l~~$ & ~~3~~  & ~~4~~ & ~~5~~ & ~~6~~ & ~~7~~ & ~~8~~\\
\hline 
$N_m$ &  0.54 & 0.51 & 0.46 & 0.39 & 0.28 & 0.06 \\
\hline
\end{tabular}
\end{table}

In the next section we will study whether the application of this $n_l$-dependent value of $N_m$ is allowing to 
get IRR-based estimates, which agree  with  the  five- and six-loop coefficients $t^M_5$ and  $t^M_6$, evaluated within both Euclidean and Minkowskian ECH-motivated approaches, and  respect   the following from the   large $\beta_0$-expansion
  sign-alternating 
behavior of their representation through   powers of $n_l$.  In order to find the answer to this problem  we should estimate the  expressions for 
$t^M_5$ and $t^M_6$-coefficients not only for the physical numbers of light   flavors  $n_l=3, 4, 5$, which corresponds 
to the cases of consideration of the charm, bottom at top-quark masses, but for unphysical values of  "light" flavors $6\leq n_l\leq 8$ as well.

\section{Numerical results and their interpretation}
\label{Numerical results}
 
We  now summarize  theoretical discussions of Sec.2-Sec.4  by comparing the 
 estimated expressions for the five and six-loop coefficients in the  the relation between pole and 
$\rm{\overline{MS}}$-scheme 
running heavy quark masses, obtained  by  the defined in   the Euclidean and Minkowskian regions ECH-motivated methods and  by  the IRR-based  asymptotic formula of Eq.(\ref{renormalon-dom}),  which is   supplemented   with the $n_l$-dependent value of the normalization factor
$N_m$. The concrete $n_l$-dependent results for the  numerical estimates of $t^M_5$ and $t^M_6$-contributions to (\ref{t^M_n}), obtained with the help of the discussed above three methods,  are given in Table 2.

\begin{table}[!h]
\centering
\caption{The estimates of $t^M_5$ and $t^M_6$-contributions by  three considered methods.}
\label{Tab-2}
\begin{tabular}{lllllll}
\hline 
$n_l$ & ~~$t^{M,\; ECH}_5$~~  & ~~$t^{M,\; ECH direct}_5$ & $t^{M,\; r-n}_5$~~ & ~~$t^{M,\; ECH}_6$~~ & ~~$t^{M,\; ECH direct}_6$ & $t^{M,\; r-n}_6$~~\\
\hline 
3 &  28435 & 26871 & 34048 &  476522 &  437146 & 829993 \\
\hline
4 &  17255 & 17499 & 22781 &  238025 &  255692 & 511245 \\
\hline
5 &  9122  & 10427 & 13882 &  90739  &  133960 & 283902 \\
\hline
6 &  3490  & 5320  & 7466  &  8412   &  57920  & 137256 \\
\hline
7 &  -127  & 1871  & 3119  &  -29701 &  15798  & 50520  \\
\hline
8 &  -2153 & -196  & 344   &  -39432 &  -2184  & 4747  \\
\hline
\end{tabular}
\end{table}

One can first  observe that for the physical values  $n_l$=3, 4, 5 the estimates for the coefficients   $t^M_5$ and $t^M_6$, 
 obtained by  two different realizations of the ECH-based technique,  are in reasonable  agreement \footnote{
It is worth emphasizing that our
 ECH-inspired results  for  $t^M_5$-coefficient at $n_l=4$ agree rather well with the   approximate value of this term,  independently  obtained 
in Ref.\cite{Mateu:2017hlz} as one of the outcomes 
 of the global fits of characteristics of the bottomonium spectrum studied in  non-relativistic QCD   up to 
N$^3$LO.}. However, {\it  it is surprising }  that 
the IRR-based approach with taken from \cite{Beneke:2016cbu} {\it four-loop values of $N_m$}
does not reproduce the {\it  supported by large $\beta_0$-approximation 
  sign-alternating $n_l$-dependence structure} \cite{Ball:1995ni}
   of the corresponding  estimated  expression for   $t^M_5$ and $t^M_6$-coefficients , obtained within both realizations of the ECH-approach.  Indeed, combining the values for $t^M_5(n_l)$ as given in three first columns of Table 2 
with the  representation of Eq.(\ref{t-n-l}),   we obtain the  following expressions:
\begin{subequations}
\begin{align}
\label{t^MECH_5}
t^{M, \; ECH}_5&=2.5n^4_l-136n^3_l+2912n^2_l-26976n_l+86620~, \\
\label{t^MECHdirect_5}
t^{M, \; ECH direct}_5&=1.2n^4_l-77n^3_l+1959n^2_l-20445n_l+72557~, \\
\label{t5ren}
t^{M, \; r-n}_5&=-22n^4_l+416n^3_l-1669n^2_l-11116n_l+72972~.
\end{align}
\end{subequations}
The similar  {\it surprising  feature} is observed at the six-loop level, namely:  
\begin{subequations}
\begin{align}
\label{t^MECH_6}
t^{M, \; ECH}_6&=-4.9n^5_l+352n^4_l-9708n^3_l+131176n^2_l-855342n_l+2096737~, \\
\label{t^MECHdirect_6}
t^{M, \; ECH direct}_6&=-2.2n^5_l+148n^4_l-4561n^3_l+71653n^2_l-538498n_l+1519440~, \\
\label{t6rn}
t^{M,  \; r-n}_6&=99n^5_l-2903n^4_l+30109n^3_l-99563n^2_l-305378n_l+2040263~.
\end{align}
\end{subequations}
This   paradox of application of the IRR asymptotic formula of Eq.(\ref{renormalon-dom}) 
is not clear to us.

Let us now consider  the asymptotic structure of the relation between pole and running masses of real  heavy quarks. One can see from Table 2 that for 
$n_l=3, 4, 5$ three methods of estimates of high-order corrections give comparable values  for the five-loop coefficients, while  the six-loop large  coefficients,  
estimated by  the  ECH-motivated approaches,  are lower than the existing  asymptotic IRR-renormalon predictions by the factor 2 only. Since for rather approximate estimate we do not consider this difference seriously, we will use these estimated by three methods numbers as inputs of our numerical studies. We fix the values of the the running masses of $c$, 
$b$ and $t$-quarks following the  presented in \cite{Kataev:2018gle} considerations as  $\overline{m}_c(\overline{m}^2_c)=1.275\; {\rm{GeV}}$,  $\overline{m}_b(\overline{m}^2_b)=4.180 \; \rm{GeV}$, $\overline{m}_t(\overline{m}^2_t)=164.3 \; \rm{GeV}$ and take 
the following    values of the $\rm{\overline{MS}}$-scheme strong coupling constant, normalized at these running masses, viz $\alpha_s(\overline{m}^2_c)=0.3947$, $\alpha_s(\overline{m}^2_b)=0.2256$,  $\alpha_s(\overline{m}^2_t)=0.1085$. Taking into account the  known results of direct diagram calculations \cite{Tarrach:1980up, Gray:1990yh, Avdeev:1997sz,   Fleischer:1998dw, Melnikov:2000qh, Chetyrkin:1999qi, Marquard:2016dcn} and using the  data, presented in Table 2, we find that within the both ECH-motivated methods and the IRR-based approach the asymptotic PT expressions for the  pole masses  of charm, bottom and top-quarks has the following form:
\begin{subequations}
\begin{align}
\label{Mc}
\frac{M_c}{1~\rm{GeV}}&\approx  1.275+0.214+0.208+0.295+0.541 \\ \nonumber &+\bigg\{\underbrace{1.135+2.389}_{\text{ECH}};~ \underbrace{1.072+2.192}_{\text{ECH direct}}; ~ \underbrace{1.359+4.162}_{\text{IRR,\; $N_m=0.54$}}\bigg\}~,  \\ 
\label{Mb}
\frac{M_b}{1~\rm{GeV}}&\approx 4.180+0.400+0.200+0.146+0.137 \\ \nonumber
&+\bigg\{\underbrace{0.137+0.137}_{\text{ECH}};~ \underbrace{0.140+0.147}_{\text{ECH direct}};~ \underbrace{0.182+0.293}_{\text{IRR,\; $N_m=0.51$}}\bigg\}~,\\
\label{Mt}
\frac{M_t}{1~\rm{GeV}}&\approx  164.300+7.566+1.614+0.498+0.196 \\ \nonumber
&+\bigg\{\underbrace{0.074+0.025}_{\text{ECH}};~ \underbrace{0.084+0.037}_{\text{ECH direct}};~ \underbrace{0.112+0.079}_{\text{IRR, \; $N_m=0.46$}}\bigg\}~.
\end{align}
\end{subequations}
Based on these results we conclude that five-loop corrections to pole mass of charm-quark are rather close in all three considered estimate methods. However, the six-loop corrections, predicted with help of the IRR technique, differ significantly from the ones, obtained by the both ECH procedures. With reference to the $b$-quark pole mass the situation is more interesting. Indeed, the ECH-motivated method that takes into account the transition from the Euclidean to Minkowskian regions demonstrates output to some kind of plateau (four, five and six-loop corrections coincide), whereas the direct ECH and the IRR approaches indicate the growth of these corrections. These facts testify to the unconditional manifestation of the asymptoticity of the corresponding PT series for bottom-quark starting with five-loop order. For case of $t$-quark all three considered estimate procedures outline the decrease of the five and six-loop corrections. This means that the asymptotic structure of this PT series is not yet manifesting itself at these levels. Therefore the conception of pole mass of top-quark can be safely used even at the $\mathcal{O}(a^6_s)$ level. 

\section{Conclusion}
We apply three approximate methods for estimation of the five and six-loop corrections to the $\rm{\overline{MS}}$-on-shell heavy quark mass relation, namely two ECH-motivated methods, defined in the Euclidean and Minkowskian regions correspondingly, and the infrared renormalon based approach. By means of these methods we determine flavor dependence of the considered contributions in the $\mathcal{O}(a^5_s)$ and 
$\mathcal{O}(a^6_s)$ orders. Wherein the IRR-based technique with normalization $N_m$-factor, taken in the four-loop approximation, does not give questionable $n_l$-dependent  results while the both ECH approaches predict not only close values of the corresponding coefficients but reproduce the sign-alternating structure of these corrections in expansion in powers of massless flavors. The numerical studies of all estimate procedures indicate the growth of the five and six-loop corrections to the pole mass of charm-quark. Whereas the ECH Euclidean method for $b$-quark pole mass leads to effect of plateau and the rest two methods outline also the increase of these corrections. In the case of $t$-quark the asymptotic nature of the corresponding PT series is not observed even at six-loop level. Therefore the concept of the pole mass of top-quark is applicable up to 6 order of PT for sure.

\section{Acknowledgments}
This contribution is based on the talks presented by one of us (V.M.) at the  XXth International Seminar on High Energy Physics ``Quarks-2018'' 
(27 May - 2 June, Valday) and at the VII International Workshop   ``Calculation for Modern and Future Colliders (CALC-2018)''  (22 July-1 August, JINR, Dubna).
We would like to thank organizers of both these events  for creating inspiring atmosphere for scientific discussions. 
The authors wish to thank  K.G. Chetyrkin, L. Dudko  and S. Moch for useful discussions at ``Quarks-2018''  Seminar and 
A.G. Grozin for valuable  comments, made after the talk at CALC-2018 Workshop. We are also grateful to V.A.Braun  for  helpful discussions.
The work of A.K.  is supported in part  by the  Russian Science Foundation grant No. 14-22-00161, while the  
 work of V.M.  and his stay in Valday was  supported by the Russian Science Foundation grant No. 16-12-10151.

\end{document}